

\magnification\magstep1
\def\ref#1{$^{[#1]}$}
\parskip = 6 pt
\def\pagenumber{\footline={\hss\tenrm\folio\hss}}

\input amssym.def      
\input amssym
\def\Hop{{\Bbb H}} 
\def\P{{\Bbb P}}   
\def\Rop{{\Bbb R}} 
\def\Zop{{\Bbb Z}} 
\def\half{{1\over 2}}

\def\bphi{ {\overline \phi}}

\def\d{\delta}

\def\inv{^{-1}}
\def\tu{\tilde u}

\def\tv{\tilde v}
\def\ag{{\alpha}}
\def\z{\zeta}
\def\bg{{\beta}}

\def\t#1,#2,{\{#1\otimes #2\}}
\def\o{\otimes}

\def\dt{{\cdot}}

\def\M{{\cal M}}
\def\p{\partial}
\def\B{{\cal B}}
\def\A{{\cal A}}

\def\ld{\lambda}

\def\sect#1{\vskip 12pt \leftline{\bf #1} \vskip 1pt}

\def\Ea{E_{\alpha}}
\def\E-a{E_{-\alpha}}

\def\da{^{\dagger}}
\def\tr{\hbox{tr}}
\def\ld{\lambda}

\def\tu{\tilde u}
\def\tv{\tilde v}
\def\ld{\lambda}

\def\tu{\tilde u}
\def\tv{\tilde v}
\def\bphi{ {\overline \phi}}
\def\o{\otimes}
\def\half{{1\over 2}}
\def\a{{\cal O}}

\def\hc{{\hat C}}
\def\dt{\cdot}
\def\ag{\alpha}
\def\da{^{\dagger}}
\def\tr{\hbox{tr}}
\def\norm{\matrix {\times \cr \times \cr}}
\def\BPZ{1}
\def\KZ{2}
\def\MS{3}
\def\MR{4}
\def\Schroer{5}
\def\Rehren{6}
\def\AGS{7}
\def\TK{8}
\def\FGK{9}
\def\FADD{10}
\def\CGHOS{11}
\def\Alek{12}
\def\MSch{13}
\def\Garb{14}
\def\Drinf{15}
\def\FKS{16}
\def\RESH{17}
\def\KR{18}
\def\RW{19}
\def\GO{20}
\def\Stegun{21}
\def\MG{22}
\def\GNOS{23}
\def\CG{24}

\nopagenumbers

\rightline{DAMTP-94-42}
\rightline{hep-th/9407108}
\vskip 20pt

\centerline{\bf QUANTISATION OF THE SU(N) WZW MODEL AT LEVEL $k$}
\vskip 50 pt

\centerline {\bf Meifang Chu and Peter Goddard}
\vskip 18 pt
\centerline {\it D.A.M.T.P}
\vskip 1pt
\centerline{\it University of Cambridge}
\vskip 1pt
\centerline{\it Silver Street, Cambridge, CB3 9EW}
\vskip 1pt
\centerline{\it United Kingdom}
\vskip 50 pt

\centerline {\bf ABSTRACT}
\vskip 6 pt
{\rightskip=18 true mm \leftskip=18 true mm

The quantisation of the Wess-Zumino-Witten model on a circle
is discussed in the case of $SU(N)$ at level $k$.
The quantum commutation of the chiral vertex operators is
described by an exchange relation with a braiding matrix, $Q$.
Using quantum consistency conditions,
the braiding matrix is found explicitly in the fundamental
representation. This matrix is shown to be related to the
Racah matrix for $U_t(SL(N))$. From calculating the four-point
functions with the Knizhnik-Zamolodchikov equations,
the deformation parameter $t$ is shown to be
$t=\exp({i\pi /(k+N)})$ when the level $k\ge 2$.
For $k=1$, there are two possible types of braiding,
$t=\exp({i\pi /(1+N)})$ or $t=\exp(i\pi)$. In the latter case,
the chiral vertex operators are constructed explicitly
by extending the free field realisation a la Frenkel-Kac and Segal.
This construction gives an explicit description of how
to chirally factorise the $SU(N)_{k=1}$ WZW model.

\noindent}

\vskip 50pt
\centerline{18 July, 1994}
\vfill\eject
\pagenumber

\sect{Introduction}

Rational conformal field theory\ref{\BPZ} has been
very useful and instructive for many aspects of two-dimensional physics.
This is largely due to the nice properties that follows
from its symmetry algebras\ref{\BPZ,\KZ}.
The generally accepted definition of
rational conformal field theory\ref{\MS} is that
the symmetry algebras are formed by two
commuting chiral algebras, ${\cal A}\times {\bar{\cal A}}$,
and each algebra contains a copy of Virasoro algebra which ensures
the conformal symmetry. In general, it also contains
other chiral algebras such as the Kac-Moody algebras
in the case of Wess-Zumino-Witten models.
Consequently, the Hilbert space can be decomposed into
a {\it finite} number of highest weight representations of
${\cal A}\times {\bar{\cal A}}$.
$$
{\cal H}= \bigoplus_{\lambda} H_{\lambda} \o
{\overline H}_{\overline \lambda}
\eqno(1)
$$

In the Euclidean formulation, there is a state/vertex operator
correspondence which says that the highest weight state in
$H_\lambda \o {\overline H}_{\overline \lambda}$
can be obtained by applying the corresponding primary field
at the origin to the vacuum,
$\Phi^{\lambda,{\overline \lambda}}(0,0)|0\rangle$.
Under conformal transformations, the behaviour of
each primary field is determined by its conformal weight
$(\Delta_\lambda, \Delta_{\overline \lambda})$.
One important consequence of these properties is that
one can calculate all the correlation functions
in the theory from those of the primary fields using
the Ward identities derived from the two conserved chiral
currents. Since the two chiral algebras commute,
these correlation functions are decomposed into a sum of
products of holomorphic and antiholomorphic
functions, which are called the {\it conformal blocks}\ref{\BPZ}.
This suggests that the primary field
$\Phi^{\lambda,{\overline \lambda}}(z,{\overline z})$ is also
decomposed into a sum of products of holomorphic primary
fields, $U^\lambda_{a}(z)$, and antiholomorphic primary fields,
$V^{\overline \lambda}_{\overline a} ({\overline z})$,
$$
\Phi^{\lambda,{\overline \lambda}}(z,{\overline z})= \sum_{a,{\overline a}}
C^{\lambda,{\overline \lambda}}_{a,{\overline a}} U^\lambda_a(z)
V^{\overline \lambda}_{\overline a} (\overline z).
\eqno(2)
$$

{}From this point of view, conformal blocks are formed by
the correlation functions of the chiral primary fields and
one should be able to derive the braiding and fusion
properties\ref{\MS} of the conformal blocks from the local properties
of these operators. In particular, they are nonlocal
operators and obey braiding relations of the following type,
$$\eqalign{
& U(z)\o U(w)=\left( 1\o U(w)\right) \left( U(z)\o 1 \right)
Q (z,w), \cr
& V({\overline z})\o V({\overline w})= {\overline Q}
({\overline z},{\overline w}) \left( 1\o V({\overline w})\right)
\left( V({\overline z})\o 1 \right). \cr}
\eqno(3)
$$
They generate the entire field content of the chiral models
and the chiral Hilbert space can be determined completely from them.
The chiral primary fields satisfying these descriptions
are called the {\it chiral vertex operators}\ref{{\MS}-{\TK}}.
(In particular, we are refering to
the {\it verter-chiral vertex operators} discussed in [{\MR}].)

It is therefore natural to ask how one can
formulate the chiral conformal field theory from
these chiral vertex operators such that their correlation functions
reproduce the conformal blocks.
Such a formulation is especially desirable from a
phenomenological point of view when one discusses the
compactification of string theories.
Whilst there has been a number of studies\ref{\MS - \CGHOS}
indicating that a quantum group related symmetry acts
in the chiral theory, a description of precisely
how the factorisation occurs and how the physical
states in the chiral Hilbert space
transform under the quantum group is lacking.

This paper is a continuation of the investigation
of this problem using the approach of canonical
quantisation begun in [{\CGHOS}] and briefly reviewed below.
The aim is to quantise the chiral primary fields in the
Wess-Zumino-Witten (WZW) models associated with a group $G$ and
compactified on a circle.
As explained in [{\FADD-\Alek}], these chiral primary fields
obey a quantum exchange relation given by a
braiding matrix, $Q$. An explicit solution for $Q$ was
obtained for $SU(2)$. In this paper, we solve for the braiding
matrix in the fundamental representation
for the model associated with $SU(N)$ at level $k$.
Because the braiding matrix can be related to the Racah matrix
for the quantum group $U_t(SL(N))$, this implies indirectly
that $U_t(SL(N))$ is acting on the chiral Hilbert space.
Instead of using the conformal scaling argument in [{\CGHOS}],
we show that the quantum value of the deformation parameter $t$
can be determined using the Knizhnik-Zamolodchikov
(KZ) equations. For $k\ge 2$, our results agree with the usual
observation that $t=\exp( i\pi/(k+N))$.
In the level-one case, we find that $U_t(SL(N))$ with
$t=e^{i\pi}$ also describes the additional symmetry in the
quantised chiral theory. Notice that when the
deformation parameter is a root of unity, the tensor
products of the irreducible representations of $U_t(SL(N))$ are
truncated to contain only the integrable representations.
These tensor products can not be made co-associative for all
irreducible representations\ref{\MSch,\Garb}.
Consequently, the corresponding symmetry algebra should be a
quasi quantum group, i.e. a quasi-Hopf algebra\ref{\Drinf}.

For the simplest models with the level $k=1$,
the chiral symmetry algebras,
i.e. level-one Kac-Moody algebras,
has an unitary realisation in terms of free scalar fields a la
Frenkel-Kac and Segal\ref{\FKS}. In this paper,
we extend this construction to give the chiral vertex operators
such that their local properties can be examined directly.
This provides an explicit description of the chiral
Hilbert space acted by two symmetry algebras,
$SU(N)_{k=1}$ Kac-Moody algebra and the
quantum group $U_{-1}(SU(N))$ (co-associative when $t=-1$).
This construction not only exemplifies
the results obtained from the canonical quantisation,
but it also provides a useful insight for formulating the
corresponding chiral WZW models.

\sect{Canonical Structures of the WZW Models}

Let us first briefly review the Poisson structures of the WZW
models derived in [{\CGHOS}].
Consider the defining field $g(\tau,x)$ which takes values in a
compact simple Lie group $G$ of rank $r$
and $x$ is the spatial coordinate
for a circle of length $2\pi$ such that
$g(\tau,x+ 2\pi )=g(\tau,x)$.
Let $\psi$ be the longest root and
$\{ t^a \} \equiv \{ H^i,E^\alpha \}$ denote a basis for
the generators of the Lie algebra, ${\cal G}$.
We shall normalise these generators so that
$Tr(t^a t^b)={\cal N} \delta^{ab}$.
The action of the WZW model has an infra-red fixed point and it
can be written as
$$
\A[g] = -{k\psi^2\over16\pi{\cal N}}
\left\{ \int_\M \tr\left(g^{-1}\partial_\mu g g^{-1}\partial^\mu
g\right)d^2x - {2\over 3}\int_\B \epsilon^{\lambda\mu\nu}
\tr\left({\hat g}^{-1}\partial_\lambda {\hat g}
{\hat g}^{-1}\partial_\mu {\hat g}{\hat g}^{-1}
\partial_\nu {\hat g} \right){d^3x} \right\},
\eqno(4)
$$
where $\B$ is a three-dimensional manifold whose
boundary is the cylinder ${\cal M}$ and ${\hat g}(\tau,x,y) \in G$
is defined on $\B$ such that it maps to $g(\tau,x)$ at $y=0$.
In order for the action to be well defined,
$k$ must be an integer so that $e^{\A}$ is single valued.

The equations of motion, i.e. the Euler-Lagrange equations obtained
from (4), can be written in terms of two chirally conserved currents:
$$
 \partial_+(g^{-1}\partial_-g)=0, \qquad
 \partial_-(\partial_+ g g^{-1})=0, \qquad
x^\pm \equiv \tau \pm x.
\eqno(5)
$$
In the gauge fixed approach\ref{\CGHOS}, a general
solution of the equations of motion can be written in terms of
the periodic coset variables $\tu \in LG/H$ and
$\tv \in H\backslash LG$:
$$
g(x,t)=\tu(x^+) e^{iq_\nu \cdot H} e^{i2(x^++x^-)\nu\cdot H} \tv(x^-).
\eqno(6)
$$

Substituting the parameterisation of (6) into the symplectic two
form, we can invert it and obtain the Poisson brackets.
In terms of the new chiral variables,
$$\eqalign{
& u(x^+)\equiv \tu(x^+) e^{i2x^+\nu\cdot H} e^{iq_\nu \cdot H}, \cr
& v(x^-)\equiv e^{iq_\nu \cdot H} e^{i2x^-\nu \cdot H} \tv(x^-),\cr}
\eqno(7)
$$
the Poisson brackets are listed as follows.
$$\eqalign{
&\{q^i_\nu ,\nu^j\} = \half \beta \delta^{ij}, \qquad
\{\tu_1(x^+),\tv_2(x^-) \}=0,\qquad
\{ \tu(x^+) , \nu^j \} =0, \qquad \{ \tv(x^-) , \nu^j \} =0, \cr
& \{u_1(x^+),v_2(y^-) \}= (x^+-y^-)\beta u_1(x^+) \Hop v_2(y^-)
+\{ \tu_1(x^+), iq_\nu  H_2 \} e^{i2x^+\nu H_1}
e^{iq_\nu H_1} v_2(y^-) \cr
&\hbox{\hskip 7cm} + u_1(x^+) e^{iq_\nu  H_2}
e^{i2y^-\nu H_2} \{ iq_\nu  H_1, \tv_2(y^-) \}, \cr
& \{ u_1(x^+), u_2(y^+) \} = u_1(x^+) u_2(y^+) \quad r (x^+-y^+), \cr
& \{ v_1(x^-), v_2(y^-) \} = {\overline r} (x^--y^-) \quad
v_1(x^-) v_2(y^-), \cr}
\eqno(8)
$$
where $\Hop \equiv H_1 \cdot H_2 = \sum_{j=1}^r H^j\o H^j$
and $\beta={4\pi/\psi^2 k}$.
The classical r-matrices in the last two brackets of (8) are
$$\eqalign{
& r (x^+) = {\bg\over 2}\eta(x^+)\Hop +i{\beta\over 2}\sum_\ag
{1\over \sin(\ag\dt\nu)}e^{-i\ag\dt\nu\eta(x^+)}E^\ag\o
E^{-\ag}, \cr
&{\overline r} (x^-) = {\bg\over 2}\eta(x^-)\Hop -i
{\beta\over 2}\sum_\ag {1\over \sin(\ag\dt\nu)}
e^{i\ag\dt\nu\eta(x^-)}E^\ag\o E^{-\ag}, \cr}
\eqno(9)
$$
with $\eta(x)=2[x/2\pi]+1$ and $[y]$ denotes the maximal
integer less than $y$. Notice that the Poisson
bracket, $\{ u_1,v_2 \}$ in (8), is different from the
one given in Eq. (36) of [{\CGHOS}] because
$\{ \tu, q_\nu \}$ and $\{ \tv, q_\nu \}$ do not vanish in general.
They can be made to vanish in a natural way for $\tu$ and
$\tv$ at the identity but will be present away from it.
Their precise forms will depend on the way the coset representations
$\tu$ and $\tv$ are chosen. We have not specified them
explicitly and they are not needed
for the discussion in the rest of this paper because
each chiral sector can be quantised separately.
The fact that the Poisson bracket in (36) of [{\CGHOS}] was
not consistent was drawn to our attention by G. Papadopoulos.
One can easily verify that the Poisson brackets in (8)
satisfy the Jacobi's identities.

Classically, we can use these Poisson brackets to verify
that $u$ ($v$) transforms as a primary field of the
right-moving (left-moving) Kac-Moody algebra, i.e.
$$
\{J^a_+(x^+),u(y^+)\}=it^a u(y^+)\d(x^+-y^+), \qquad
 \{J^a_-(x^-),v(y^-)\}=i v(y^-)t^a\d(x^- - y^-).
\eqno(10)
$$
The Kac-Moody currents can be written in terms of the
chiral group elements $u(x^+)$ and $v(x^-)$ as
$$
J_+(x^+)= (ik/4\pi)\p_+u u\inv, \qquad
J_-(x^-)= (ik/4\pi)v\inv \p_-v.
\eqno(11)
$$

\sect{Quantisation and the Braiding Matrix}

Let us proceed to quantise these Poisson brackets.
The first four brackets in (8) can simply
be replaced by Dirac commutators,
$$
[q^i_\nu , \nu^j ]= i {\beta \hbar \over 2} \delta^{ij}, \qquad
[\tu_1(x^+), \tv_2(y^-) ]=0,\qquad [ \tu (x^+), \nu^j ] =0,
\qquad [ \tv (x^-), \nu^j ] =0.
\eqno(12)
$$
The quantum correction to the constant $\beta \hbar$ will be
determined later from consistency conditions.
Following [{\CGHOS}], we quantise the last two brackets in (8)
by the following exchange relation,
$$
\eqalignno{
& u_1(x^+) u_2(y^+) = u_2(y^+) u_1(x^+) Q(x^+-y^+), &(13a)\cr
& v_1(x^-) v_2(y^-) = Q^{-1}(x^--y^-) v_2(y^-) v_1(x^-).&(13b)\cr}
$$
where $Q(x)$ is the braiding matrix which can be determined
from the consistency of quantisation as follows.
$$\eqalignno{
&\hbox{{\it Unitarity}}, \hbox{\hskip 2.8cm} Q\da (x) = Q^{*T}(x) =
Q^{-1}(x) &(14a) \cr
&\hbox{{\it Antisymmetry property}},\qquad Q(-x) = \P Q(x)^{-1}\P\
&(14b) \cr
&\hbox{{\it Monodromy property}},\hbox{\hskip 1cm}
Q(x+ 2\pi )=(M(\nu)^{-1}\o {\cal I})Q(x)(M(\nu_2)\o {\cal I})
& (14c)\cr
&\hbox{\hskip 5.7cm} =({\cal I}\o M(\nu_1))Q(x)({\cal I}\o
M(\nu)^{-1})\,&(14d)\cr
&\hbox{{\it Classical limit}},\hbox{\hskip 2cm}
Q(x)={\cal I}\o{\cal I}+ i\hbar r (x) + O(\hbar^{2})\,
&(14e) \cr
&\hbox{{\it Locality condition}}, \hbox{\hskip 1.5cm}
[ Q(x), e^{iq_\nu \cdot (H_1 + H_2)}]=0 & (14f) \cr
&\hbox{{\it Jacobi identities}},\qquad{\rm with}\quad
x_{ij}\equiv x_i-x_j, \cr
&\qquad  Q_{23}(\nu_1,x_{23}) Q_{13}(\nu,x_{13})
Q_{12}(\nu_3,x_{12}) =Q_{12}(\nu,x_{12})Q_{13}(\nu_2,x_{13})
Q_{23}(\nu,x_{23}), &(14g) \cr}
$$
The subscript of $\nu$ denotes a shift,
e.g. $\nu_2 = \nu + \half  \hbar \beta H_2$,
due to the commutation relation,
$$
[u(x), \nu]=- \half \beta \hbar u(x) \cdot H.
\eqno(15)
$$
These six requirements are fairly obvious except the
locality condition, (14f), which is added to ensure
that two $g$'s at a space-like separation commute with each other.

In the following, we solve for the braiding matrix in the
N-dimensional fundamental representation for $SU(N)$ at
level $k$. Let $\Phi$ denotes the space of roots and
${\vec \ld}_a$ for $a=1,2,..N$ be the
weights in the fundamental representation whose highest
weight is $\lambda_1$.
The normalisation of these vectors is chosen to be
${\vec \ld}_a \cdot {\vec \ld}_b=\delta_{ab}-1/N$.
Let $e_{ab}$ denote a $N\times N$ matrix which has only
one non-zero element, $1$, located at the $a^{\rm th}$
row and the $b^{\rm th}$ column. Then,
the Cartan-Weyl basis is given by
${\vec H}= \sum_{a=1}^N {\vec \ld}_a e_{aa}$
and the step-operator $E_{\alpha_{ab}}=e_{ab}$
for ${\vec \ag}_{ab}= {\vec \ld}_a-{\vec \ld}_b \in \Phi$.
With this normalisation, ${\cal N}=1$ and
the length of the longest root is $\psi^2=2$.

In order to solve for $Q$ in (14),
we exponentiate the classical r-matrix in (9) and replace
its coefficients by their quantum values:
namely, $\chi$, $\theta$, and $\gamma$.
$$\eqalign{
Q(x)= &\exp \left\{ i\chi\eta(x) \Hop - \sum_{\alpha \in \Phi} \theta
(\alpha\dt\nu) e^{-i\gamma_\alpha \eta(x)} E_\ag \o
E_{-\ag} \right\}, \cr
 =& t^{({N-1\over N})\eta(x)} \left( 1\o 1-
\sum_{\alpha\in\Phi} \Ea \E-a \o \E-a \Ea \right) \cr
&+ t^{-{1\over N}\eta(x)} \sum_{\alpha\in\Phi} \cos \theta
(\ag\dt\nu ) \Ea \E-a \o \E-a \Ea \cr
& - t^{-{1\over N}\eta(x)}\sum_{\alpha\in\Phi}
e^{-i\gamma_\alpha  \eta (x)} \sin \theta (\ag\dt\nu )
\Ea \o \E-a, \cr}\qquad \quad t\equiv e^{i\chi}.
\eqno(16)
$$
Without specifying the values of these coefficients,
(16) already satisfies most of the requirements in (14)
except the monodromy property and the Jacobi's identities.
These two properties impose constraints on the coefficients
in $Q$ and they can be solved as follows.
$$
\gamma_\alpha =\ag \dt \nu, \qquad
\sin \theta(\ag\dt\nu) = {\sin (\chi) \over \sin (\ag \dt \nu)},
\qquad \chi={1\over2}\beta\hbar
\longrightarrow {\pi\over k} + O(\hbar).
\eqno(17)
$$
Therefore, the braiding matrix is determined completely
up to the braiding parameter $t$.
The quantum value of this parameter will be determined in the
next section when we study the monodromy properties of the
four-point functions by solving the Knizhnik-Zamolodchikov equations.

Notice that this braiding matrix can be related to the
Racah matrix of $U_t(SL(N))$ and consequently indicates
that $U_t(SL(N))$ is an additional symmetry of the chiral theory.
This identification comes from the following observation.
{}From the conformal scaling argument given in the appendix,
we know that $\alpha \cdot \nu / \chi \in \Zop$
because the monodromy takes values of
$\nu / \chi = \Lambda + \delta$ for some
highest weight $\Lambda$ and $\delta$ denotes one half of
the sum of all positive roots. Subsequently, when $t \ne \pm 1$,
the braiding matrix $Q$ in (16) is a $N^2 \times N^2$
matrix related to the deformed Racah matrix\ref{\RESH-\RW}
for $U_t(SL(N))$ up to some normalisation:
$$
Q_{mn,m'n'}(\nu) \propto \delta (\lambda_m + \lambda_n -
\lambda_{m'} - \lambda_{n'})
\left\{ \matrix {\lambda_1 & \Lambda & \Lambda -\lambda_{n'} \cr
\lambda_1 & \Lambda - \lambda_m - \lambda_n
& \Lambda -\lambda_{m} \cr} \right\}^{\rm RW}_{q=t^2}.
\eqno(18)
$$
This identification is valid only when all the entries
in the bracket are highest weights.
In order to avoid confusion,
we have denoted the quantum group by $t=q^\half$
rather than $q$ such as in [{\RESH,\KR}].
In the limit when $t^2 =1$, $Q$ can be related to
the undeformed Racah matrix of $SU(N)$.

For example, when $N=2$ and $t\ne \pm 1$, we can
substitute $\sqrt{2}\nu/\chi=(2j+1)$ for $j\ge 1$ and obtain
the following $4\times 4$ matrix in terms of the
deformed-Racah matrix of $U_t(SL(2))$,
$$\eqalign{
& Q_{mn,m'n'}(\nu)
= \delta_{m+n,m'+n'} (-1)^{\ell+L-j-J} t^{ (C_j+C_J-C_\ell-C_L)}
\left\{ \matrix {\half & j & j-{n'\over 2} \cr
\half & j-{m+n \over 2} & j-{m\over 2} \cr}
\right\}^{\rm RW}_{q=t^2}, \cr
&{\rm with}\quad C_j=j(j+1),\quad \ell = j- {n'\over 2},
\quad L = j-{m\over 2}, \quad J = j- {m+n \over 2}; \forall
m,n,n'=\pm 1. \cr}
\eqno(19)
$$
In the limit when $t=e^{i\pi}$, this gives the
Racah($\{ 6j \}$)-matrix for $SU(2)$ according to
$$
Q_{mn,m'n'} ( \nu ) =\delta_{m+n,m'+n'} (-1)^{mn\over 2}
(-1)^{n-n' \over 2} \left\{ \matrix {\half & j & j-{n'\over 2} \cr
\half & j-{m+n \over 2} & j-{m\over 2} \cr} \right\}.
\eqno(20)
$$
As explained in the next section, the quantum value of $\chi$
is typically given by $\pi$ divided by a positive integer. This
raises a concern that (16) and (17) become singular
when $\alpha \cdot \nu / \pi \in \Zop$.
However, for the WZW models which we are studying,
only the primary fields in the integrable representations\ref{\GO}
contribute to correlation functions,
i.e their highest weights satisfy $0 < \lambda^0 \cdot \psi < k+1$.
Restricted to these representations only\ref{\MR},
the monodromy momentum takes the value of
$\nu / \chi =\lambda^0 + \delta$ and it satisfies
$0< \alpha \cdot \nu / \pi < 1$ when $\chi=\pi/(k+N)$.
When the level $k=1$, the coefficient
$\sin \theta (\alpha \cdot \nu)$ in (17) becomes a finite integer
in the limit of $\chi=\pi$ for $\alpha \cdot \nu / \pi \in \Zop$.
Therefore, the braiding matrix in (16) is regular for
the physical representations considered
in the chiral $SU(N)_k$ WZW models.

\sect{Braiding Matrix and Monodromy Matrix}

In order to determine the quantum value of the braiding parameter,
$t=e^{i\chi}$, one can follow the conformal scaling argument
given in [{\CGHOS}]. We give the result of that argument for $SU(N)$
in the appendix. The drawback of this argument is
that it does not apply to the case when $k=1$.
In this section, we give a new derivation of the braiding parameter
based on quantum consistency, i.e. the
Knizhnik-Zamolodchikov (KZ) equations.
Since $Q$ in (16) describes the braiding relation of two chiral
primary fields in the fundamental representation, we shall
consider the four-point functions in the
representations: $\Lambda_1, \Lambda,\Lambda,\Lambda_4$; namely,
the two vertex operators in the middle are in the fundamental
representation, $\Lambda$. Let us denote the weights in these
representations by $a_j$, $b_j$
and define the following four-point functions,
$$
F_{a_1 a_2 a_3 a_4}^{b_1 b_2 b_3 b_4} (z_1,z_2,z_3,z_4) \equiv
\langle 0| U_{a_1 b_1}^{\Lambda_1}(z_1)U_{a_2 b_2}^{\Lambda}(z_2)
U_{a_3 b_3}^{\Lambda}(z_3)U_{a_4 b_4}^{\Lambda_4}(z_4)|0\rangle.
\eqno(21)
$$
By exchanging the vertex operators at $z_2$ and $z_3$
according to (13a), we find the following braiding
relation between the four-point functions,
$$
F_{a_1 a_2 a_3 a_4}^{b_1 b_2 b_3 b_4} (z_1,z_2,z_3,z_4)=\sum_{r_2,r_3}
F_{a_1 a_3 a_2 a_4}^{b_1 r_3 r_2 b_4} (z_1,z_3,z_2,z_4)
Q_{r_2 r_3,b_2 b_3}\left( {z_2 \over z_3},t \right).
\eqno(22)
$$

The main idea is to solve for the four-point functions
using the Ward-identities and then determine the braiding
parameter $t$ from (22).
As shown in [{\KZ}], these four-point functions
satisfy the following set of differential equations,
$$
\eqalignno{
& \sum_{j=1}^4 z_j^n \left( z_j \partial_{z_j} + (n+1) \Delta_j
\right) F^{\{ b \} } =0; \qquad n=-1,0,1 & (23a)\cr
& \left( \sum_{j=1}^4 t^\alpha_j \right)
F^{\{ b \} } =0, & (23b) \cr
& (k+N) \partial_{z_i} F^{\{ b \}} = \sum_{j=1,j\ne i}^4
{t_i \cdot t_j \over z_i - z_j } F^{\{ b \}}, & (23c)\cr}
$$
where $\{ t_j \}$ denote the $SU(N)$ generators in $\Lambda_j$
representation and $\Delta_j$ denotes the conformal weight
of the $j$-th vertex. These weights are given by
their highest weight vectors according to
$\Delta_j=\langle \Lambda_j,\Lambda_j +2\delta \rangle /2(k+N)$.
For example, $\Delta ( \Lambda )=(N^2-1)/2N(k+N)$
for the fundamental representation.

Before solving the KZ equations, let us first recall that,
following from (23a), the correlation functions depend only on the
coordinate differences, $z_{ij}\equiv z_i-z_j$. Once a pre-factor
independent of the indices $\{ a \}$ and $\{ b \}$ has been
extracted, the rest will depend only on the cross-ratio,
$ y\equiv (z_{12}z_{34})/ (z_{13} z_{24})$,
$$\eqalign{
& F_{\{ a \} }^{\{ b \}} (z_1,z_2,z_3,z_4)
= \left( \prod_{i<j} z_{ij}^{-\gamma_{ij}} \right)
f_{\{ a \} }^{\{ b \}} (y), \qquad \cr
{\rm where} \qquad \qquad &
\gamma_{12}=\gamma_{13}=0, \qquad \gamma_{14}=2\Delta_1, \qquad
\gamma_{23}=\Delta_1+\Delta_2 + \Delta_3 - \Delta_4, \cr
& \gamma_{24}=-\Delta_1+\Delta_2 - \Delta_3 + \Delta_4,
\qquad \gamma_{34}=-\Delta_1-\Delta_2 + \Delta_3 + \Delta_4 .\cr}
\eqno(24)
$$
Thus, omitting the $\{ b \}$ indices for the moment,
we can rewrite the KZ equation as follows,
$$
(k+N)\partial_y f (y) = \left( {t_1 \cdot t_2 \over y}
+ {t_1 \cdot (t_2 + t_3) \over 1-y} \right) f (y).
\eqno(25)
$$

For example, let us first consider the case when
$\Lambda_1=\Lambda_4={\overline \Lambda}$ conjugate to
the fundamental representation $\Lambda$.
The weights in $\Lambda$ are denoted by
$m\equiv \lambda_m$ for $m=1,2,...N$ and those
in ${\overline \Lambda}$ by ${\overline m}\equiv -\lambda_m$.
Eq. (23b) implies that $f_{\{ a \} }^{\{ b \}}$ is an isotropic
tensor on the lower indices ${\{ a \} }$ and so can be written
in terms of a basis of such tensors; there are two independent
ones in this case,
$$
f_{{\overline m},n,l,{\overline k}} (y) \quad = \quad
\delta_{m,n}\delta_{l,k} f_1 (y)+ \delta_{m,l}\delta_{n,k} f_2 (y).
\eqno(26)
$$
In this case, the KZ equation in (25) becomes a set of differential
equations in terms of the two scalars,
$$
(k+N) \partial_y \left( \matrix { f_1 \cr f_2 \cr}
\right) (y) = \left\{ {1 \over y}
\left( \matrix{{1-N^2 \over N} & -1 \cr 0& {1\over N} \cr} \right)
- {1 \over 1-y}  \left(
\matrix{{N^2-2\over N} & 1 \cr 1& {N^2-2 \over N} \cr}\right) \right\}
\left( \matrix {f_1 \cr f_2 \cr}\right) (y).
\eqno(27)
$$
This equation can be rewritten as second-order differential
equations whose solutions are hypergeometric functions.
In order to relate to the braiding relation in (22),
it is sufficient to consider the solutions in
the limit of $z_{12} \rightarrow 0 $ or $y \rightarrow 0$.
There are two independent sets of solutions:
$$\eqalign{
&f_1^- (y)= y^{-2\Delta} (1-y)^{N^2+N-2\over N(k+N)}
F\left( {1\over k+N},{k+1\over k+N},{k\over k+N},y \right), \cr
&f_2^- (y)=-{1\over k} y^{1-2\Delta} (1-y)^{N^2+N-2\over N(k+N)}
F \left( {1+{1\over k+N}},{k+1\over k+N},2-{N\over k+N},
y \right),\cr}
\eqno (28a)
$$
and
$$\eqalign{
&f_1^+ (y)= y^{1\over N(k+N)} (1-y)^{N^2+N-2\over N(k+N)}
F \left( 1+{1\over k+N},{N+1\over k+N},1+{N\over k+N},y \right),  \cr
&f_2^+ (y)= -N y^{1\over N(k+N)} (1-y)^{N^2+N-2\over N(k+N)}
F \left( {1\over k+N},{N+1\over k+N},{N\over k+N},y \right).\cr}
\eqno(28b)
$$

It is easy to check that these solutions are the same as those
in Eq. (4.10) of [{\KZ}] where the KZ equation was solved
in terms of the cross ratio
$ x\equiv y/(y-1)=(z_{12}z_{34})/ (z_{14} z_{32})$.
The reason why we have used $y$ in this paper
is because the monodromy properties
of these solutions are easier to study on the $y$-plane.
We shall explain the details in the next paragraph.
Meanwhile, one can understand why there are
these two sets of solutions as follows. According to the
operator product expansions (OPE) of the chiral primary fields,
$$
U^{\Delta_1}(z_1) U^{\Delta_2}(z_2)  \sim  \sum_{\Delta_3}
(z_1-z_2)^{\Delta_3  - \Delta_1-\Delta_2 }
U^{\Delta_3}(z_2) \left( 1 + O(z_1-z_2) \right),
\eqno(29)
$$
we know that $U^{\Delta_3}$ corresponds to the
intermediate states contributing to the four-point functions.
Thus, the singular behaviours of the solutions in (28)
near $z_{12} \rightarrow 0$, together with the prefactor
$\left( \prod_{i<j} z_{ij}^{-\gamma_{ij}} \right)$,
imply that the first set of solutions, $f^-(y)$, is
contributed from the intermediate states in the singlet
representation of $SU(N)_k$ and the second set of solutions,
$f^+(y)$, is contributed from those in the
adjoint representation. These are the two
intermediate representations expected from the
fusion rule\ref{\MS}.

We can now determine the braiding parameter in (22) from the
monodromy properties of the four-point functions.
In general, the $SU(N)$-invariant correlation function is
a linear combination of these two solutions:
$$
F^{ \{ b \} }= \left( \prod_{i<j} z_{ij}^{-\gamma_{ij}} \right)
\sum_{\rho =\pm} Y_\rho^{ \{ b \} } f^\rho (y).
\eqno(30)
$$
We can denote the solutions of (28) graphically
and write the braiding relation in (22) as
$$
\sum_{\rho=\pm}\quad\quad {\Biggr\rangle^{\hbox{\hskip -22pt
{2,m}}}_{\hbox{\hskip -22pt {1,${\overline m}$}}}} {\hskip -1pt}
{\hbox to 20pt{\leaders\hrule\hfill}}
\rho {\hbox to 20pt{\leaders\hrule\hfill}}{\hskip -1pt}
\Biggl\langle^{\hbox{3,n}}_{\hbox{4,${\overline n}$}}
Y_\rho^{\{ b \}} = \sum_{{\rho'=\pm}\atop r_2,r_3} \quad \quad
{\Biggr\rangle^{\hbox{\hskip -22pt {3,n}}}_{\hbox{\hskip -22pt
{1,${\overline m}$}}}}{\hskip -1pt}
{\hbox to 20pt{\leaders\hrule\hfill}}
\rho'{\hbox to 20pt {\leaders\hrule\hfill}}{\hskip -1pt}
\Biggl\langle^{\hbox{2,m}}_{\hbox{4,${\overline n}$}}
Y_{\rho'}^{b_1 r_3 r_2 b_4} Q_{r_2 r_3,b_2 b_3}.
\eqno(31)
$$
When we exchange $z_2$ with $z_3$, the cross-ratio
$y$ becomes $1/y$. Thus, we need to relate the
hypergeometric functions of $y$ to those of $1/y$.
Since $0<y<1$ for $0 <z_3 < z_2 < \infty$,
we can choose the cut of these functions to be the negative real axis.
Using the properties of the hypergeometric functions\ref{\Stegun},
we find that $f_1 (y)$ and $f_2 (1/y)$ are linearly
related by a connection matrix $\cal K$ as follows.
$$\eqalignno{
& f_1^\rho (y) = e^{i\pi{N^2-1 \over N(k+N)}} \sum_{\mu=\pm}
f_2^\mu ( {1\over y} ) {\cal K}_\mu^\rho, &(32a) \cr
& {\cal K}= e^{-i{\pi\over N(k+N)}} \left( \matrix{
{e^{i{N\pi\over k+N}} \sin({\pi\over k+N})\over
\sin({N\pi\over k+N})} & k ({1-N^2 \over N}) {
\Gamma^2 (1+{N\over k+N}) \over \Gamma(1+{N+1\over k+N})
\Gamma(1+{N-1\over k+N})} \cr
 {-1\over N} {\Gamma^2(1-{N\over k+N})\over
\Gamma(1-{N+1\over k+N})\Gamma(1-{N-1\over k+N})} &
- { e^{-i{N\pi \over k+N}} \sin({\pi\over k+N})\over
\sin({N\pi\over k+N})} \cr} \right). &(32b) \cr}
$$

In deriving for ${\cal K}$, we have implicitly
chosen a contour of analytic continuation
by specifying $ -y = e^{i\pi\eta}y $
with $\eta \equiv \eta({\rm arg}(z_2 / z_3))=1$.
One can check that the following results apply to the other
range of $z_2 / z_3$.
Notice that the normalisation of $f_1$ determines
the normalisation of $f_2$ and the definition
of the connection matrix depends on this normalisation. However,
the eigenvalues of the connection matrix are not affected.
{}From (28), (30) and (32), we find that the braiding relation
in (22) will hold provided
$$
\sum_{\mu=\pm} {\cal K}_\rho^\mu Y_{\mu}^{b_1 b_2 b_3 b_4}
=\sum_{r_2,r_3=1}^N  Y_{\rho}^{b_1 r_3 r_2 b_4}
Q_{r_2 r_3, b_2 b_3}(\nu^0_4,t).
\eqno(33)
$$
Here, $\nu^0_4$ denotes the value of the momentum $\nu$ acting
on the vacuum with a shift due to the vertex operator
at $z_4$, see (15). This condition requires that
$\P Q$ and ${\cal K}$ have at least one common eigenvalue.
$\P$ is the permutation matrix which interchanges
the left two indices of $Q$.
{}From (16), $\P Q$ has two distinct eigenvalues,
$t^{N-1 \over N}$ of multiplicities $\half N(N+1)$ and
$-t^{-{N+1\over N}}$ of multiplicities $\half N(N-1)$.
While, the connection matrix ${\cal K}$ has two
eigenvalues: $e^{i\pi(N-1)/N(k+N)}$ and $-e^{-i\pi(1+N)/N(k+N)}$.
This would give totally four possible values for $t$.
In general, we do not expect $Y^{\{ b \} }$ to vanish
when $b_2=b_3$. Then, only two of the four possibilities for
$t$ should be considered.
$$\eqalignno{
&t=e^{i\pi/(k+N)}, & (34a) \cr
{\rm or}\qquad & t^{{N-1 \over N}}=(-1)
e^{-i\pi {N+1 \over N(k+N)} }. & (34b) \cr}
$$
Further consideration is needed in order to determine
which of these two gives the correct braiding parameter.
In the following, we will discuss the case for $k=1$ and
$k\ge 2$ separately.

When the level $k=1$, it is known\ref{\GO} that only the chiral
primary fields in the minimal representations can contribute
to the correlation functions. Therefore, the adjoint solution
$f^+$ in (28b) is not allowed, i.e. $Y_+^{ \{ b \} } =0$ in (30).
Substituting the singlet solution $f^-$ in (28a) into the
braiding relation in (31), we find that
$$
e^{i2\pi\Delta_2} Y_-^{b_1 b_2 b_3 b_4} = \sum_{r_2,r_3=1}^N
 Y_-^{b_1 r_3 r_2 b_4} Q_{r_2 r_3, b_2 b_3}(\nu^0_4,t).
\eqno(35)
$$
If the diagonal coefficients $Y^{b_1 r r b_4}\ne 0$,
then the braiding parameter must be $t=e^{i\pi}$ as in (34b).
However, if one can impose an extra symmetry
in the chiral theory\ref{\MG} so that $Y^{b_1 r r b_4} = 0$,
then (34a) can also be a solution of (35).
To summarise, for the $SU(N)_{k=1}$ WZW model,
$$\eqalignno{
& t= e^{i{\pi\over 1+N}} \qquad\quad {\rm only}\quad{\rm if}\quad
Y^{b_1 r r b_4} = 0, & (36a)\cr
&t=e^{i\pi}\qquad\qquad {\rm if}\qquad\qquad Y^{b_1 r r b_4} \ne 0.
& (36b) \cr}
$$

This seems to indicate that there may be two different ways of
factorising the level-one $SU(N)$ WZW model by implementing
the two different internal symmetries.
Previous treatments\ref{\MS-\AGS,\Garb} have proposed the quasi-quantum
group $U_t(SL(N))$ with $t=e^{i{\pi\over 1+N}}$
in (36a) to be the internal symmetry although
explicit implementations have not been given.
On the other hand, the possibility of $U_{-1}(SL(N))$ symmetry
has not been considered, and we shall show that
it is realised explicitly by the free field construction
of the chiral vertex operators in the next section.

Let us now consider what happens when $k \ge 2$.
Since both solutions in the singlet and adjoint representations
are present, we have to study other types of four-point functions
in order to determine the quantum value for $t$.
In the following, we shall consider the four-point functions in
representations: $\Lambda_1,\Lambda,\Lambda,\Lambda$, with
$\Lambda_1$ conjugates to the totally symmetric representation
of ${{N(N+1)(N+2)/6}}$ dimensions.
The conformal weight for $\Lambda_1$ is given by
$\Delta_1= 3(N+3)(N-1)/2N(k+N)$. In this case, there is only
one isotropic tensor $C$ and the correlation function
can be written as $C$ multiplied by a scalar $S^{\{ b \} }$,
$$
F_{a_1,a_2,a_3,a_4}^{\{ b \} }(\{ z \} ) = C_{a_1,a_2,a_3,a_4}
S^{\{ b \} }(\{ z \} ), \qquad \left( \sum_{j=1}^4 t_j \right) C =0 .
\eqno(37)
$$
Using the fact that $\Lambda_2=\Lambda_3=
\Lambda_4=\Lambda$ and $ C_{a_1,a_2,a_3,a_4}$ is totally
symmetric in the last three indices, we have
$$
\left( t_1 \cdot t_2 \right) C =
- \left( {N^2-1 \over N}+ 2 {N-1 \over N} \right) C.
\eqno(38)
$$
Similarly for $\left( t_1 \cdot t_3 \right)$ on $C$,
we can solve for the scalar $S^{\{ b \} }$ from (25):
$$
S^{\{ b \}}=Y^{\{ b \}} \left( z_{24}z_{34}
\right)^{\Delta_1-\Delta_N} z_{14}^{-2\Delta_1}
z_{23}^{-\Delta_1-\Delta_N} \left( y(1-y)^2
\right)^{-{(N-1)(N+3)\over N(k+N) }},
\eqno(39)
$$
where $Y^{\{ b \}}$ is some normalisation constant.
Substituting (37) and (39) into the braiding relation in (22),
we find that the braiding parameter is given by (34a),
$$
t=\exp ( i {\pi\over k+N} )\qquad {\rm for}\qquad k\ge 2.
\eqno(40)
$$
Therefore, when $k \ge 2$, this result agrees with the previous
observations in [{\MS-\CGHOS}] that the additional symmetry
acting on the chiral sectors of $SU(N)_k$ WZW model is the
quasi-quantum group $U_t(SL(N))$ with the parameter $t$ given
in (40).

For $N=2$, our results on the monodromy properties
of the four-point functions agree with
those of Tsuchiya and Kanie\ref{\TK} for both $k=1$ and $k\ge 2$,
but it should be noted that we have used a different
basis to solve the KZ equations in order to be able to
relate the connection matrix ${\cal K}$ with
the braiding matrix $Q$ as in (33).
This distinction is significant when we are considering the
construction of the chiral parts $U(x^+)$, $V(x^-)$ of the group
element $g(x^+,x^-)$, rather than just chiral primary fields.
A relevant observation is that
the (quasi-)quantum group should act on $U(x^+)$ from the right-hand side
while the Kac-Moody current acts from the left-hand side.
In this way, the (quasi-)quantum group appears only in the chiral
theory and it plays the role of an {\it internal} symmetry of the
original WZW model.

In this section, we have completed solving for
the braiding matrix specifying the exchange relation of
two chiral vertex operators in the fundamental
representation. In principle, this allows us to determine
also the braiding relations involving
chiral vertex operators in other representations.
This is because the OPE's of the chiral vertex operators
in the fundamental representation generate the rest of
the chiral vertex operators. Indirectly, this result
also reveals the symmetry structures of the Hilbert
space of the chiral model. In order to make this
more precise, we discuss in the next section
the explicit construction of the
chiral vertex operators in terms of free fields
for the level-one $SU(N)$ WZW model.

\sect{Free Field Construction of the Chiral Vertex Operators at
Level One}

There are two stages in our construction: the first stage is to
obtain the chiral primary fields by extending the construction of
Frenkel-Kac and Segal\ref{\FKS} for the affine algebras. It turns
out that these chiral primary fields obey a diagonal braiding
relation. The second stage is to determine the zero-modes
in order to complete the construction of the chiral vertex
operators defined in (13a) and (16).

Let us briefly review the free field realisation of the
level-one Kac-Moody algebras in [{\FKS}].
Further details can be obtained in [{\GO}].
The main advantage of this construction is that
unitarity is manifest.
In the Cartan-Weyl basis, this affine algebra can be written
as the following with the structure constants
$\epsilon(\alpha,\beta)$ normalised to be $\pm 1$.
$$\eqalign{
&[J^i_m,J^j_n]=m\delta^{ij}\delta_{m,-n} \hbox{\hskip 3cm}
i,j=1,2,..N-1.\cr
&[J^i_m,J^\alpha_n]=\alpha^i J^\alpha_{m+n} \cr
&[J^\alpha_m, J^\beta_n]=\cases{
\epsilon (\alpha,\beta) J^{\alpha+\beta}_n
    & if $\alpha +\beta$ is a root \cr
\sum_{j=1}^r \alpha^j J^j_{m+n} +m\delta_{m,-n}
    & if $\alpha +\beta =0$ \cr
0   & otherwise. \cr} \cr}
\eqno(41)
$$
Let us introduce $N-1$ right-moving scalar fields $\phi^j (z)$ which
is expanded on a complex plane of $z=e^{ix^+}$,
$$
\phi^j(z)= q_\phi^j - i p_\phi^j \ln z + i \sum_{n\ne 0, \in \Zop}
{1\over n} \phi^j_n z^{-n},
\eqno(42a)
$$
where
$$\eqalign{
& \phi^{j\dagger}_n=\phi^j_{-n},\qquad \quad [\phi^i_m,\phi^j_n]=
m\delta_{m,-n}\delta^{ij}, \qquad j=1,2,...N-1; \cr
&  p_\phi^{j\dagger}=p_\phi^j ,\qquad\qquad
[q_\phi^i,p_\phi^j]=i\hbar \delta^{ij}. \cr}
\eqno(42b)
$$
For each root $\beta$,
we define a normal ordered vertex operator,
$$\eqalign{
\a^{\beta}(z) & \equiv z^{\beta^2/2 }
\norm \exp(i\beta \cdot \phi(z)) \norm \cr
&\equiv \sum_{n \in Z} z^{-n} \a^{\beta}_n \cr
&=  z^{\beta^2/2} \exp({-\beta \cdot \sum_{n<0}{1\over n}
\phi_n z^{-n}}) \exp({i\beta \cdot q +\beta \cdot p \ln z}) \exp(
{-\beta\cdot \sum_{n>0}{1\over n} \phi_n z^{-n}}). \cr}
\eqno(43)
$$
The product of two such operators has a short distance
expansion obtained by normal ordering,
$$
\a^\alpha(z) \a^\beta(w) = (z-w)^{\alpha \cdot \beta}
\norm \a^\alpha(z) \a^\beta(w) \norm \qquad {\rm for} \quad |z| > |w|.
\eqno(44)
$$
This allows us to evaluate the following
contour integral by integrating $z$ over the contour ${\cal C}_w$
which winds around $w$,
$$
\int_{{\cal C}_0} {dw\over 2\pi i} \int_{{\cal C}_w}
{dz\over 2\pi i} (z-w)^{\alpha \cdot \beta} z^{m}w^{n-1}
\norm \a^\alpha(z) \a^\beta(w) \norm = \a^{\alpha}_m \a^{\beta}_n -
(-1)^{\alpha \cdot \beta} \a^{\beta}_n \a^{\alpha}_m.
\eqno(45)
$$

The right-hand side of (45) almost gives the commutator of
$\a^\alpha$ and $\a^\beta$ except for the extra phase
$(-1)^{\alpha \cdot \beta}$ in front of the second term.
In order to remove this phase, we introduce an operator $\hc_\alpha$
defined on the root lattice $\Lambda_R$ such that they
obey the following cocycle condition,
$$\eqalign{
& \hc_\alpha \hc_\beta = S(\alpha,\beta) \hc_\beta \hc_\alpha
= \epsilon(\alpha,\beta) \hc_{\alpha+\beta}, \cr
{\rm with}\qquad & S(\alpha,\beta)=e^{i\pi \alpha \cdot \beta}, \quad
\forall \alpha,\beta \in \Lambda_R. \cr}
\eqno(46)
$$
The structure constant $\epsilon (\alpha,\beta)$
and the symmetry factor $S(\alpha,\beta)$ have to satisfy
the following consistency requirements,
$$\eqalignno{
&\epsilon (\alpha,\beta) \epsilon (\alpha+\beta,\gamma)
=\epsilon(\alpha,\beta+\gamma) \epsilon(\beta,\gamma), & (47a) \cr
&\epsilon(\alpha,\beta)=S(\alpha,\beta)
\epsilon(\beta,\alpha), & (47b)\cr
&S(\alpha,\beta)S(\gamma,\beta)=S(\alpha+\gamma,\beta),& (47c) \cr
&S(\alpha,\beta)=S^{-1}(\beta,\alpha). & (47d) \cr}
$$
These conditions determine $\hc$ up to some gauge
transformation.\ref{\GO,\GNOS} We will discuss this gauge freedom
in more details later on. In general, without introducing
new degrees of freedom, one can construct $\hc$
as functions of the zero modes $p,q$ of the free scalar fields.
{}From (42)-(47), the current generators can be realised
as follows.\ref{\FKS}
$$\eqalign{
&J^{\alpha}(z) = \a^\alpha (z) e^{-i\alpha\cdot q_\phi} \hc_\alpha,
\qquad \forall \quad \alpha \quad \hbox{\rm is a root},\cr
&J^j (z) = i z \partial_z \phi^j(z), \quad \qquad
\qquad j=1,2,..N-1. \cr}
\eqno(48)
$$
Conformal symmetry is generated by the Virasoro
generators which can be obtained from the affine currents
according to the Sugawara construction,
$$
L(z)={1\over 2(1+N)}\norm \sum_{a=1}^{N^2-1}
J^a(z) J^a(z) \norm = {1\over 2}\norm (iz\partial_z \phi)^2 \norm
\eqno(49)
$$

The left-moving sector can be constructed similarly with
$N-1$ left-moving scalar fields which depend on
${\overline z}\equiv e^{ix^-}$,
$$
\bphi({\overline z}) = {\overline q}_\bphi -i{\overline p}_\bphi
\ln {\overline z} +i \sum_{n \ne 0} {1\over n}
\bphi_n {\overline z}^{-n}
\eqno(50)
$$
Similarly, the left-moving Kac-Moody generators and
the Virasoro generators are given by
$$\eqalign{
&{\overline J}^{\alpha}({\overline z}) = {\overline z}^{\beta^2/2}
 \norm \exp(i\beta \cdot \bphi({\overline z})) \norm
({\overline z}) e^{-i\alpha\cdot {\overline q}_\bphi }
{\overline C}_{\alpha}, \qquad \forall
 \quad \alpha \quad \hbox{{\rm is a root}},\cr
&{\overline J}^j ({\overline z}) = i {\overline z}
{\overline \partial} \bphi^j({\overline z}),
{\hbox{\hskip 3cm}} \qquad j=1,2,...N-1.\cr
&{\overline L}({\overline z})={1\over  2(1+N)}\norm
 \sum_{a=1}^{N^2-1} {\overline J}^a({\overline z})
{\overline J}^a({\overline z}) \norm = {1\over 2} \norm
(i{\overline z}{\overline \partial}
{\overline \phi})^2 \norm \cr}
\eqno(51)
$$

In order to extend these results
to construct the chiral primary fields,
it is necessary to introduce a finite number
of additional degrees of freedom into the chiral sector.
The reason is much clearer if we first explain the construction
of the non-chiral primary fields $g(z,{\overline z})$.
To begin with, we consider $g$ as a $N\times N$ matrix
in the fundamental representation. Its conformal weight is given by
$\Delta_N={\overline \Delta_N}=(N^2-1)/2N(1+N)$. Let $t_N$'s denote
the generators of $SU(N)$ in the fundamental representation.
Then, $g$ transforms under the Kac-Moody algebras and the
Virasoro algebras according to
$$\eqalignno{
&[J^a_m, g (z,{\overline z})]=z^m t_N^a g (z,{\overline z}),
\quad\quad [L_m,g (z,{\overline z})]= z^m
(z \partial_z +(m+1) \Delta_N ) g (z,{\overline z}), &(52a) \cr
& [{\overline J}^a_m, g (z,{\overline z})]= {\overline z}^m
g (z,{\overline z}) t_N^a, \quad\quad [{\overline L}_m,g_{ab}
(z,{\overline z})]={\overline z}^m ( {\overline z}
\partial_{\overline z} +(m+1){\overline \Delta_N} )
g (z,{\overline z}). & (52b)\cr}
$$

{}From the vertex operators in (43) and the contour
integrals in (45), we observe that the $z$ and
${\overline z}$-dependent parts of $g_{rs}(z,{\overline z})$
should be $\a_{\lambda_r}(z)$ and ${\overline \a}_{\lambda_s}
({\overline z})$ where $\lambda$'s are
the weights in the fundamental representation.
Then, the solution for $g(z,{\overline z})$ in (52) has
the following form,
$$
g_{rs}(z,{\overline z})=
\a_{\lambda_r}(z){\overline \a}_{\lambda_s}
({\overline z}) C_{(\lambda_r,\lambda_s)}
(p_\phi,{\overline p}_\bphi ).
\eqno(53)
$$
The non-trivial part is to construct the cocycle on the direct
product of two weight lattices,
$$
\hc_{\vec \lambda} \equiv e^{i(\lambda \cdot q_\phi
+{\overline \lambda} \cdot {\overline q}_\bphi )}
C_{\vec \lambda}(p_\phi ,{\overline p}_\bphi),
\qquad \forall \quad  {\vec \lambda} \equiv
(\lambda,{\overline \lambda})
\in \Lambda_W \times \Lambda_W.
\eqno(54)
$$
On this enlarged lattice, the cocycles for the Kac-Moody
generators correspond to $\hc_{(\alpha,0)}$ and
$\hc_{(0,\alpha)}$ respectively. Then, the primary fields will
have the correct commutators with the current
generators provided that the symmetry factors are
given by
$$
S({\vec \mu},{\vec \nu})= e^{i\pi(\mu \cdot \nu - {\overline \mu} \cdot
{\overline \nu})},\qquad \forall \quad {\vec \mu}, {\vec \nu}
\in \Lambda_W \times\Lambda_W.
\eqno(55)
$$
One can easily check that (55) satisfy the
consistency conditions in (47c) and (47d) on the new lattice
$\Lambda_W \times \Lambda_W$.

Once the symmetry factor is given,
we can determined $\epsilon({\vec \mu},{\vec \nu})$
using (47b) up to the following gauge transformation
with $ t({\vec \mu})=t(\mu) {\overline t}({\overline \mu})$,
$$
\hc_{\vec \mu} \longrightarrow \quad t({\vec \mu})
\hc_{\vec \mu} ; \qquad
\epsilon({\vec \mu},{\vec \nu}) \longrightarrow \quad
{t({\vec \mu})t({\vec \nu})\over t({\vec \mu}+{\vec \nu})}
\epsilon({\vec \mu},{\vec \nu}).
\eqno(56)
$$
Subsequently, the cocycles can be constructed as functions
of the zero-modes, ${\vec p}\equiv (p_\phi,{\overline p}_\bphi)$.
$$
\hc_{\vec \lambda}  \equiv e^{i{\vec \lambda}\cdot {\vec q}}
\epsilon ({\vec \lambda},{\vec p}).
\eqno(57)
$$
The gauge freedom in (56) allows us to normalise the symmetry factor
$S({\vec \mu},{\vec \mu})=1$. This implies that the left and
the right weights, $\mu$ and ${\overline \mu}$ are actually in the
``diagonal" subspace,
$$
\Lambda^\pm_D \equiv \bigcup_{\lambda=0,minimal}
\Lambda_\lambda \times \Lambda_{\pm \lambda}.
\eqno(58)
$$
In the following discussion, we shall mainly consider the
cocycles in $\Lambda_D^+$ since it is compatible with
the original Hilbert space of the WZW models.

Thus, we have completed the construction of the primary field
$g(z,{\overline z})$ in terms of free scalar fields. This allows us
to verify directly that $g$ is local in the sense that
$$
g_{rs}(z,{\overline z}) g_{r's'}(w,{\overline w})= g_{r's'}(w,{\overline w})
g_{rs}(z,{\overline z}), \qquad {\rm for} \quad |z|>|w|.
\eqno(59)
$$
Since we consider the WZW models compactified on a circle,
$g(z,{\overline z})$ is periodic and this
amounts to impose an extra constraint on the left-moving and
the right-moving momenta,
$$
\lambda_r \cdot p_\phi - \lambda_s \cdot {\overline p}_\bphi
\in Z, \qquad \forall \lambda_r,\lambda_s \in [\lambda].
\eqno(60)
$$
This is manifestly satisfied when $( p_\phi,{\overline p}_\bphi)$
takes values in $\Lambda_D^+$.
This can be understood as the gauge-fixed approach discussed
in [{\CGHOS}].

Now, it is fairly clear that the left-moving and the
right-moving degrees of freedom in the primary field
of (53) are nicely separated except for the cocycles which
depend on the zero-modes of both the left-moving and the
right-moving scalar fields.
One way to separate these degrees of freedom is
to enlarge the phase space by introducing, in each chiral
sector, $(N-1)$ new pairs of harmonic operators,
$[q^l_\pm, p^j_\pm]=i\delta^{lj}$.
These new degrees of freedom are used to replace
the zero-modes of the opposite sector in the cocycles.
To be more explicit, we write down the right-moving
primary field in the fundamental representation as
$$
{\hat U}_{rs}(z) = \norm e^{i\lambda_r \cdot \phi(z)} \norm
e^{i\lambda_s \cdot q_-} C_{(\lambda_r,\lambda_s)}(p,p_-).
\eqno(61)
$$
The left-moving primary field $V({\overline z})$
can be constructed in a similar fashion using $q_+,p_+$.

The advantage of having such an explicit construction
is that we can calculate the correlation functions and
verify the KZ equation (23c) directly
using free scalar fields and cocycles. Also,
the local properties of the chiral primary fields
can be determined directly. In fact, we find
the following braiding,
$$\eqalign{
& {\hat U}_{1}(z) {\hat U}_{2}(w)= {\hat U}_{2}(w)
{\hat U}_{1}(z) R \left( \arg (z/w) \right),
\qquad {\rm for}\quad |z|>|w|, \cr
{\rm where} \qquad  & R_{rs,r's'}(x) =
\delta_{rr'} \delta_{ss'} \exp \left\{i\pi\lambda_s \cdot
\lambda_{s'}\eta \left( x \right) \right\}. \cr}
\eqno(62)
$$
When $\eta (x)=1$, $R(x)$ can be written as
$$
R(t) = t^{-{1\over N}} \left\{ t \sum_{r=1}^N e_{rr}
\otimes e_{rr} + \sum_{r\ne s=1}^N e_{rs} \otimes e_{rs}
+ (t-t^{-1}) \sum_{r>s=1}^N e_{rs}\otimes e_{sr} \right\},
\quad t=e^{i\pi}.
\eqno(63)
$$
Thus, we can identify $e^{i{1\over N}\eta(x)}R(x)$ as the
$R$-matrix\ref{\RESH} in the fundamental representation of
$U_t(SU(N))$ with $t=e^{i\pi}$.
Note that we have replaced $U_t(SL(N))$ by
$U_t(SU(N))$ since $t=-1 \in \Rop$. This braiding parameter
agrees with the one in (36b) obtained from solving the
KZ equations.

However, the braiding relation in (62) is given by
the R-matrix rather than the braiding matrix $Q$ in (16).
In order not to alter the defining properties of chiral primary
fields, we can multiply the free-field vertex ${\hat U}(z)$
by a $N$-dimensional matrix, $A$, which
commutes with the Kac-Moody currents.
In particular, let the matrix $A$ depend on the momentum, $p_-$
and denote ${\hat A}\equiv e^{iHq_-}A(p_-)$ such that
$$
U(z) \equiv \norm e^{i\phi(z) H} \norm \left( e^{-iq_\phi H} \right)
{\hat C} \left( e^{-iq_- H} \right) {\hat A}.
\eqno(64)
$$
Then, the braiding of $U$ will be given by $Q$ provided that
${\hat A}$ satisfies the following equation,
$$
R(t) {\hat A}_1 {\hat A}_2 = {\hat A}_2 {\hat A}_1 Q (\nu,t).
\eqno(65)
$$
It is sufficient to show that (65) holds for $\eta (x)=1$
because the same result will follow automatically for other $x$.
In fact, this requirement gives the operator identity for
the IRF-Vertex transformation\ref{\MR,\FGK,\RESH,\CG}
for $U_t(SL(N))$ when ${\hat A}$ is identified as the
Wigner-matrix which relate states in different irreducible
representations of $U_t(SL(N))$.

For $U_t(SL(2))$, this operator identity
has been realised explicitly in [{\CG}]
in terms of two pairs of harmonic oscillators.
In particular, we take their result in the
limit of $[Y]_t=Y t^{Y-1}$ for $Y \in \Zop$ when $t=e^{i\pi}$
and obtain
$$
{\hat A}\equiv e^{i{\sigma_3 \over \sqrt{2}} q_-}
\left( \matrix { \cos \theta &
-e^{-i\pi {(\nu + p_-)\over \sqrt{2}}} \sin \theta \cr
e^{i\pi {(\nu + p_-)\over \sqrt{2}}} \sin \theta &
\cos \theta \cr} \right)
e^{i{\sigma_3 \over \sqrt{2}} {q_\nu}},
\qquad \sin \theta \equiv \sqrt{{\nu}-p_- \over 2{\nu}}.
\eqno(66)
$$
Notice that ${\hat A}$ depends on two pairs of harmonic
oscillators, $(q_-, p_-)$ and an additional pair $(q_\nu, \nu)$
where $\nu$ is the monodromy momentum in the
braiding matrix $Q$. These zero modes are independent degrees of freedom
in the sense that they commute with the Kac-Moody current.
We will come back to the discussion of the monodromy
in a short while.

Thus, we have completed the construction of
the chiral vertex operator, $U(z)$ in (64).
Its unitarity is manifest because both ${\hat C}$ and
${\hat A}$ are unitary.
According to [{\CG}], ${\hat A}$ transforms covariantly
under $U_{-1}(SU(2))$, i.e.
$$
\varrho \left( A_{+b}, A_{-b} \right)= \left( A_{+b}, A_{-b} \right)
(\Pi^{\half} \o id) \Delta (\varrho), \qquad \forall \quad
\varrho \in U_{-1}(SU(2)),
\eqno(67)
$$
where $\Pi^{\half}$ denotes the spin $\half$ representation
and the coproduct $\Delta (\varrho)$ can be obtained from
those of the quantum group generators,
$$
\Delta (t^{S_3})=t^{S_3}\o t^{S_3}, \qquad
\Delta (S_\pm)= t^{\mp S_3} \o S_\pm + S_\pm \o  t^{\pm S_3}.
\eqno(68)
$$
Since $t^2=1$, the symmetry algebra $U_{-1}(SU(2))$
is isomorphic to $SU(2)$. Its generators can be realised
in terms of the harmonic oscillators as
$$
S^3\equiv {1\over \sqrt{2} } p_-, \qquad S^\pm \equiv
e^{\pm i \sqrt{2}q_-} \left( t^{\mp{\nu\over \sqrt{2}}} \right)
\sqrt{ \left( {\nu\over \sqrt {2}} \right)^2
- \left( {p_- \over \sqrt{2}} \pm \half \right)^2 }.
\eqno(69)
$$
The transformation laws of (67)
can be verified directly with (66) and (69).
Moreover, the quadratic Casimir operator of the algebra
is given by $({\nu \over \sqrt{2}})^2 -{1\over 4}$.
Thus, its irreducible representations, ${\overline V}_j$,
are labelled by the eigenvalues of
$({\nu \over \sqrt{2}} - \half)=j$ for $j=0,\half,1,{3\over 2},..$
The Wigner operator ${\hat A}$, consequently the chiral vertex
operator $U$, takes a state in ${\overline V}_j$ to
another state in ${\overline V}_{j\pm \half}$
of $U_{-1}(SU(2))$. Because this algebra commutes with the chiral
affine algebras, it plays the role of an internal symmetry
in the original WZW model.

Let us now discuss the monodromy momentum, $\nu$.
{}From its definition given in (7), the chiral vertex operator has
the following monodromy,
$$
U_{rs}(ze^{i2\pi})=U_{rs}(z) e^{i2\pi \nu \cdot \lambda_s}.
\eqno(70)
$$
Comparing with the explicit construction of $U$ in (64),
we need to impose a constraint between $\nu$ and the
free-field momentum $p_\phi$ such that
$$
e^{i2\pi p_\phi \cdot \lambda_r} = e^{i2\pi \nu \cdot \lambda_s},
\qquad \forall \lambda_r,\lambda_s=\pm {1\over \sqrt{2} }.
\eqno(71)
$$
This condition is manifestly satisfied when the
spins of $U_{-1}(SU(2))$ and the spin of the
Kac-Moody representation differ by an integer.
Applying the chiral vertex operator and the Kac-Moody
current successively on the vacuum, we obtain all the states in the
chiral Hilbert space. It can be decomposed as follows.
$$
{\cal H}_R=\Bigl( H_0 \o ( \oplus_{j=0,1,2,3,..}
{\overline V}_j ) \Bigr) \bigoplus
\left( H_{\half} \o  ( \oplus_{j={\half,{3\over 2},..}}
{\overline V}_j ) \right).
\eqno(72)
$$

\sect{Hilbert Space for the Chiral $SU(N)_{k=1}$ WZW Model}

The results obtained in the last section
for the chiral vertex operators in the fundamental
representation of $SU(2)$ can be generalised to other
basic representations of $SU(N)$.
The chiral Hilbert space is decomposed as
$$
{\cal H}_R=\bigoplus_{\lambda=0,{\rm minimal}}
 H_\lambda \o \left(\oplus_{\xi \in {\tilde \Lambda}_\lambda}
{\overline V_\xi} \right),
\eqno(73)
$$
where $H_\lambda$'s are the minimal representations of
the right-moving $SU(N)_{k=1}$ Kac-Moody algebra and
${\tilde \Lambda}_\lambda$ denotes the space
of all the highest weights of $U_{-1}(SU(N))$ contained in
the weight lattice, $\Lambda_\lambda$. Then, the monodromy
constraint in (71) will hold for any two weights $\lambda_r$
and $\lambda_s$ in the same representation.

The Hilbert space of the left-moving sector can be
constructed in a similar way using the left-moving
free scalar fields and $(N-1)$ additional pairs of
harmonic oscillators.
$$
{\cal H}_L=\bigoplus_{\lambda=0,{\rm minimal}}
\left( \oplus_{\xi \in {\tilde \Lambda}_\lambda}
{V_\xi} \right) \o {\overline H_\lambda}.
\eqno(74)
$$
These new degrees of freedom do not appear in the
original WZW models and they should be
gauged away when we ``join" the left-moving and the
right-moving sectors together. In particular, we should
recover the original defining field $g(z,{\overline z})$
in (53) and also the original Hilbert space
$$
{\cal H}_{WZW} = \bigoplus_{\lambda=0,{\rm minimal}}
H_{\lambda} \o {\overline H_\lambda}.
\eqno(75)
$$
One has to make this gauging process more precise
in order to obtain the Lagrangian formulation of the
chiral WZW model. We hope to investigate this question
in the near future.

{}From the free field construction of the chiral vertex
operators, we obtain an explicit description of how to
factorise the $SU(N)_{k=1}$ WZW model into the right-moving
and the left-moving sectors. In contrast
to the previous treatments\ref{\MS-\CGHOS,\Garb}, the chiral
Hilbert space acquires an additional symmetry
given by $U_{-1}(SU(N))$ instead of $U_t(SL(N))$ with
$t=e^{i\pi/(1+N)}$. From the study of the KZ
equations, see (36), we learn that these correspond to
the only two possible braidings which describe the local
properties of the chiral vertex operators
in the level-one chiral models.
However, an explicit realisation of the quasi-quantum group
symmetry $U_t(SL(N))$ with $t=e^{i\pi/(1+N)}$ is still lacking.
When $k\ge 2$, we know from the braiding matrix that
there is a unique chiral facotrisation with
the additional symmetry given by the quasi-quantum group
$U_t(SL(N))$ with $t=\exp ( i\pi/(k+N) )$.
We are currently investigating its implementation in the
chiral WZW models of higher levels.
At the moment, the different nature of the chiral
factorisation in the level-one WZW models
remains somewhat mysterious.

\vfill\eject

\sect{Acknowledgements}
We would like to thank David Olive for many helpful
discussions in the early stages of this work. We are also
grateful to Fedor Smirnov for helpful discussions about
the IRF-Vertex relation.
This work is supported by the Science and Engineering Research
Council under the grant GR/H57929.

\sect{Appendix: Conformal Scaling and the Braiding Parameter}

In this appendix, we use the scaling argument\ref{9} to
determine the braiding parameter $t$ for the $SU(N)_k$
WZW model when $k\ge 2$. Let us denote $\z =e^{i2\pi x/l}$.
For some constant $\mu$, $\z^\mu u(\z)$ transforms
as a primary field under the Sugawara stress tensor,
$$
L={1\over2(k+N)} \sum_{a=1}^N \norm J^a J^a \norm.
\eqno(A1)
$$
The conformal weight of $u(\zeta)$ in the $[\Lambda]$ representation
is given by $h= {<\Lambda,\Lambda+ 2 \delta>\over 2(k+N)}$.
Under a scaling, $\z \mapsto s\z$, with $s=e^{2\pi i}$,
$u$ transforms according to
$$
s^{L_0} u^\Lambda (\z) s^{-L_0} = s^{h-\mu} u^\Lambda (\z)
e^{2i\nu \dt H}.
\eqno(A2)
$$
The left-hand side comes from the monodromy of the $u$-vertex.
Consider the matrix elements of (A2) between two states
$\langle \Lambda_L,\rho_L |$ and $|\Lambda_R,\rho_R \rangle$
in the highest weight representations $\Lambda_L$ and $\Lambda_R$.
We find that the non-zero elements of this matrix
for $u_{rs}$ satisfy
$$
\rho_L=\rho_R + \lambda_r, \qquad \Lambda_L=\Lambda_R+\lambda_s.
\eqno(A3)
$$

For $k>1$, we can consider $\Lambda_{L,R}$ in the range
$0< \Lambda_L \cdot \psi < k$ and $0< \Lambda_R \cdot \psi < k$.
Then, with $\nu |\Lambda_R,\rho_R \rangle=\nu_R
 |\Lambda_R,\rho_R \rangle$ and similarly for $\nu_L$,
(A2) gives the following constraints for all
$\lambda_s$ in the fundamental representation,
$$\eqalign{
&{1\over \pi} \nu_R \cdot \lambda_s -
{{(\Lambda_L-\Lambda_R, \Lambda_R+\delta)+{1\over 2}
(\Lambda_L-\Lambda_R)^2}\over {k+N}} +h - \mu \in \Zop ; \cr
&(\nu_L-\nu_R)\cdot \lambda_s -{1\over 2}\hbar \beta
\lambda_s^2 \qquad \in \quad \Zop.\cr}
\eqno(A4)
$$
By choosing $\nu_R$ in the positive Weyl
chamber and excluding the extreme representations such as
$\Lambda_L \cdot \psi = 0, k$ and $\Lambda_R \cdot \psi = 0, k$,
we find
$$\cases{
\nu_L= (\Lambda_L+ \delta) \left( {\pi\over k+N} \right) , \qquad
\nu_R= (\Lambda_R+ \delta) \left( {\pi\over k+N} \right), \cr
\nu_L-\nu_R= \half \hbar \beta (\Lambda_L-\Lambda_R). \cr}
\eqno(A5)
$$
Subsequently, we are lead to the following result:
$$
\chi= \half \hbar \beta = {\pi\over k+N},
\qquad {\rm i.e.} \qquad t \equiv e^{i\chi}=e^{i{\pi\over k+N}},
\quad {\rm for} \quad k \ge 2.
\eqno(A6)
$$

\leftline{\bf References}

\item{ [{\BPZ}]} A.A. Belavin, A.M. Polyakov, A.B. Zamolodchikov,
Nucl. Phys. {\bf B241} (1984) 333.

\item{ [{\KZ}]} V.G. Knizhnik and A.B. Zamolodchikov, Nucl. Phys.
{\bf B247} (1984) 83.

\item{ [{\MS}]} G. Moore, N. Seiberg, Comm. Math. Phys. {\bf 123}
(1989) 177; Phys. Lett. B. 212 (1988) 451.

\item{ [{\MR}]} G. Moore, N. Reshetikhin, Nucl. Phys. {\bf B328}
(1989) 557.

\item{ [{\Schroer}]} B. Schroer, Nucl. Phys. {\bf B295} (1988) 4.

\item{ [{\Rehren}]} K. -H. Rehren Commun. Math. Phys. {\bf 116}
(1988) 675.

\item{ [{\AGS}]} L. Alvarez-Gaume, C. Gomez, G. Sierra,
Phys. Lett. {\bf B220} (1989) 142.

\item{ [{\TK}]} A. Tsuchiya, Y. Kanie, Lett. Math. Phys.
{\bf 13} (1987) 303, Advanced Studies in Pure Mathematics,
{\bf 16} (1988) 297.

\item{ [{\FGK}]} G. Felder, K. Gawedzki and A. Kupiainen,
Commun. Math. Phys. {\bf 130} (1990) 1.

\item{ [{\FADD}]} L.D. Faddeev, Commun. Math. Phys. {\bf 132}
(1990) 131.

\item{ [{\CGHOS}]} M. Chu, P. Goddard, I. Halliday, D. Olive,
A. Schwimmer, Phys. Lett. {\bf B266} (1991) 71.

\item{ [{\Alek}]}A. Alekseev, S. Shatashvili,
Commun. Math. Phys. {\bf 133} (1990) 353.

\item{ [{\MSch}]} G. Mack, V. Schomerus, Nucl. Phys. {\bf B370}
(1992) 185.

\item{ [{\Garb}]} M. R. Garberdiel, ``An explicit construction
of the quantum group in chiral WZW-models", preprint
DAMTP-94-51.

\item{ [{\Drinf}]} V.G. Drinfel'd, Leningrad Math. J. {\bf 1}
(1990) 1419.

\item{ [{\FKS}]} I.B. Frenkel, V.G.Kac, Invent. Math. {\bf 62}
(1980) 23; G. Segal, Commun. Math. Phys. {\bf 80} (1981) 301.

\item{ [{\RESH}]} N. Yu. Reshetikhin, ``Quantised Universal
Enveloping Algebras, The Yang-Baxter Equation and
Invariants of Links, I, II", LOMI preprint E-4-87.

\item{ [{\KR}]} A.N. Kirillov, N. Yu. Reshetikhin,
``Infinite-dimensional Lie algebras and groups" edited by
V. G. Kac, World Scientific, Singapore, (1989) pp 285.

\item{ [{\RW}]} F. Pan, Jour. Phys. A. Math. Gen. {\bf 26}
(1993) 4621.

\item{ [{\GO}]} P. Goddard, D. Olive, Int. Jour. Mod. Phys.
{\bf A1} (1986) 303.

\item{ [{\Stegun}]} See Eq. (15.3.7) in
``Handbook of Mathematical Functions" edited by
M. Abramowitz, I. A. Stegun, Dover Publications, Inc., New York.

\item{ [{\MG}]} Private communication with M. Gaberdiel.

\item{ [{\GNOS}]} P. Goddard, W. Nahm, D. Olive, A. Schwimmer,
Commun. Math. Phys. {\bf 107} (1986) 179.

\vfill\eject

\item{ [{\CG}]} M. Chu, P. Goddard, ``Quantisation of a particle moving
on a group manifold", preprint DAMTP-94-41.

\end